\documentclass[a4paper,11pt]{article}
\usepackage{Proceeding_template/pos}
\usepackage{float}

\hypersetup{pdfauthor={Nicolas De Angelis}}

\title{Development and science perspectives of the POLAR-2 instrument: a large scale GRB polarimeter}
 \ShortTitle{POLAR-2: a large scale GRB polarimeter}

\author*{Nicolas De Angelis}
\affiliation{DPNC, University of Geneva, Switzerland}
\emailAdd{nicolas.deangelis@unige.ch}

\forColl{POLAR-2\footnote{\url{https://www.unige.ch/dpnc/polar-2}}}

\abstract{Despite several decades of multi-wavelength and multi-messenger spectral observations, Gamma-Ray Bursts (GRBs) remain one of the big mysteries of modern astrophysics. Polarization measurements are essential to gain a more clear and complete picture of the emission processes at work in these extremely powerful transient events. In this regard, a first generation of dedicated $\gamma$-ray polarimeters, POLAR and GAP, were launched into space in the last decade. After 6 months of operation, the POLAR mission detected 55 GRBs, among which 14 have been analyzed in detail, reporting a low polarization degree and a hint of a temporal evolution of the polarization angle. Starting early 2024 and based on the legacy of the POLAR results, the POLAR-2 instrument will aim to provide a catalog of high quality measurements of the energy and temporal evolution of the GRB polarization thanks to its large and efficient polarimeter. Several spectrometer modules will additionally allow to perform joint spectral and polarization analyzes. The mission is foreseen to make high precision polarization measurements of about 50 GRBs every year on board of the China Space Station (CSS). The technical design of the polarimeter modules will be discussed in detail, as well as the expected scientific performances based on the first results of the developed prototype modules.}

\FullConference{37$^{\rm{th}}$ International Cosmic Ray Conference (ICRC 2021)\\
		July 12th -- 23rd, 2021\\
		Online -- Berlin, Germany}

\graphicspath{{./Pictures/}}

\begin{document}
\maketitle

\section{Introduction}

Gamma-Ray Bursts (GRBs) are among the most energetic and bright events in the universe. Discovered in 1967 by the US Vela satellites originally sent to monitor USSR nuclear tests, they remain one of the big mystery of modern astrophysics. GRBs consist of a prompt emission in the $\gamma$-ray band, that can last from fractions of seconds to tens of minutes, followed by an afterglow spanning the full electromagnetic range from radio to TeV energies. The first important science results on these transient sources came from the BATSE experiment, which measured the location of a few thousands GRBs \cite{BATSE_catalog}, showing a uniform distribution in the sky and therefore implying that GRBs are extragalactic events. The duration of the prompt emission $T_{90}$ measured by BATSE showed that GRBs can be classified in two classes, short and long GRBs, depending on whether the prompt emission is shorter or longer than 2s. The short GRBs present a relatively hard spectra and are nowadays known to be originated by neutron star mergers, as showed by the observation of GRB170817A with a gravitational wave counterpart \cite{GW170817}. Long GRBs, which typically present softer spectra, are associated with the death of supermassive stars, called hypernovae. Despite half a century of GRB temporal and spectral measurements, the physical processes at play in these sources are still poorly known. In order to gather more information about the astrophysics of GRBs, polarization measurements are essential, since they provide two new probes into the emission models: the polarization degree (PD) and angle (PA).

\section{Polarimetric Measurements of Gamma-Ray Bursts with POLAR}

\begin{figure}[H]
\centering
\includegraphics[height=0.45\textwidth]{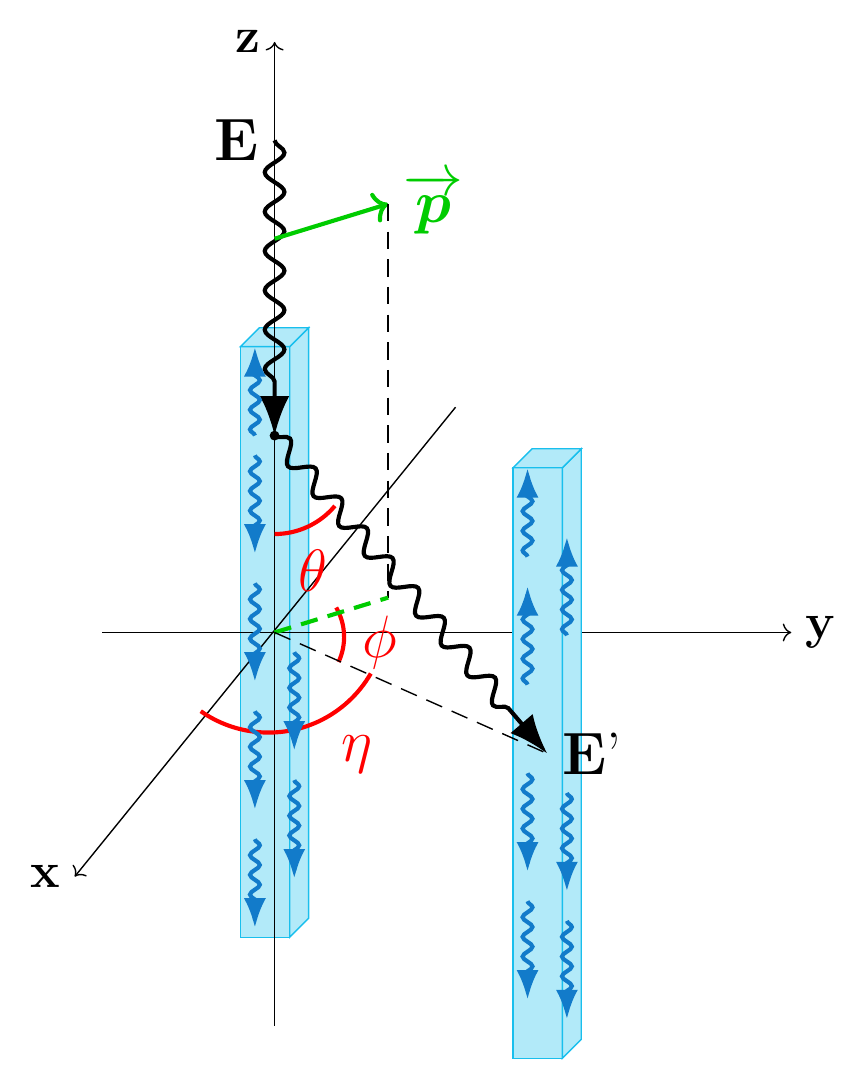}\hspace*{0.5cm}\includegraphics[height=0.45\textwidth]{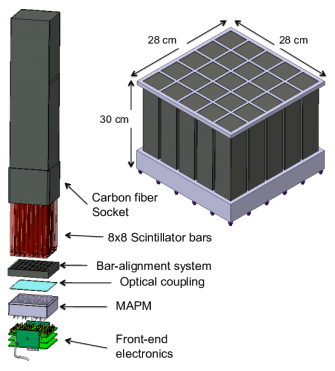}
 \caption{Illustration of the Compton scattering used for polarization measurements (left) and design of the POLAR instrument \cite{POLAR_design} (right)}
 \label{comptonscat_polar}
\end{figure}

A measurement of the polarization of a $\gamma$-ray source can be achieved by making use of the Compton scattering, ruled by the Klein-Nishina cross section. Indeed, the azimuthal scattering angle is linked to the PA of the incoming photon since this latter will preferentially scatter orthogonal to its initial polarization vector. A segmented array of scintillator bars can therefore be used in order to measure the incoming $\gamma$-rays which would scatter in a first bar and be absorbed (or scattered once again) in a second bar. The deposited energy in each elongated scintillator will then be converted into optical light, readout by optical sensors at the extremity of the bar. The relative position of the two bars allows to compute the scattering angle, whose distribution, called modulation curve, leads to the PA (less populated angle in the distribution) and PD (relative amplitude of the modulation). This measurement method is depicted in the left part of Figure \ref{comptonscat_polar} and has been used by the POLAR experiment \cite{POLAR_design}.\\

The POLAR experiment \cite{POLAR_design}, whose design is shown on the right side of Figure \ref{comptonscat_polar}, has been launched in September 2016 together with the Tiangong-2 (TG-2) Chinese space lab. It took scientific data for about 6 months and has detected 55 GRBs in the 50-500~keV energy range. The instrument consisted of an array of 40$\times$40 long scintillator bars (5.8$\times$5.8$\times$176~mm$^3$) divided into 5$\times$5 modules, all readout by Multi-Anode Photo-Multiplier Tubes (MAPMTs) and housed in carbon fiber sockets. Plastic scintillators were used for the polarimeter sensitive volume in order for the Compton scattering to be dominant on the full POLAR energy range (low-Z material).

A detailed polarization analysis of the five brighter GRBs detected by POLAR has first been published in \cite{POLAR_NatureAstronomy}, followed by a catalog paper with polarization and joint spectral analysis (with Fermi-GBM and Swift-BAT data) of 14 GRBs \cite{POLAR_catalog}. This latter seems to indicate a lowly polarized emission from most of these GRBs, as discussed in \cite{MK_ICRC2021}. Time-resolved analysis showed a hint for continuous evolution of the PA with time that washes out the PD on time-integrated analysis for several GRBs \cite{POLAR_time_resolved}, but the lack of statistics does not allow to exclude a discrete single change of the PA. Some non GRB related analyses were also performed using the POLAR data and are still ongoing, especially spectral and polarization analyses of the Crab pulsar as discussed in \cite{POLAR_Crab_spectral} and more recently in \cite{HC_Crab_ICRC2021}.\\

A conclusion that can be drawn from the POLAR experiment is that polarimetric measurements with higher statistics are needed in order to perform proper time and energy resolved analyses and to be able to disentangle between the various emission models proposed by the theoretical community. This motivation led to the development of a new generation large scale $\gamma$-ray polarimeter: the POLAR-2 instrument.

\section{Development of the POLAR-2 Large Scale GRB Polarimeter}

A successor of the POLAR experiment, called POLAR-2, has been in development for about 2 years and will be launched to the China Space Station (CSS) in early 2024. Proposed by a Swiss (UniGe), Polish (NCBJ), Chinese (IHEP), and German (MPE) collaboration, it aims to provide high quality polarization measurements of GRBs. The overall instrument design as well as the polarimeter modules development and testing are described hereafter.

\subsection{Overall Instrument Design}

\begin{figure}[H]
\centering
\includegraphics[height=0.45\textwidth]{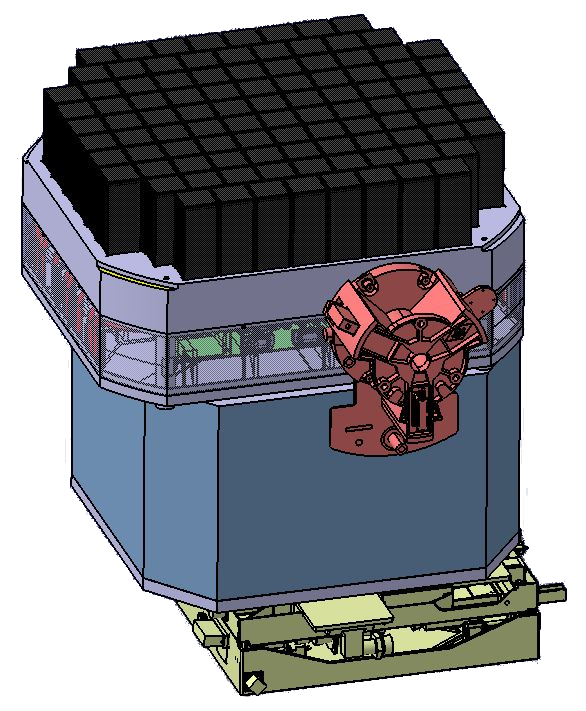}\hspace*{0.5cm}\includegraphics[height=0.45\textwidth]{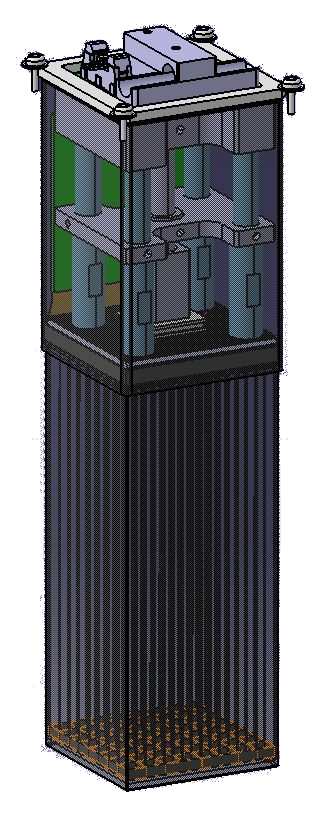}\hspace*{0.5cm}\includegraphics[height=0.45\textwidth]{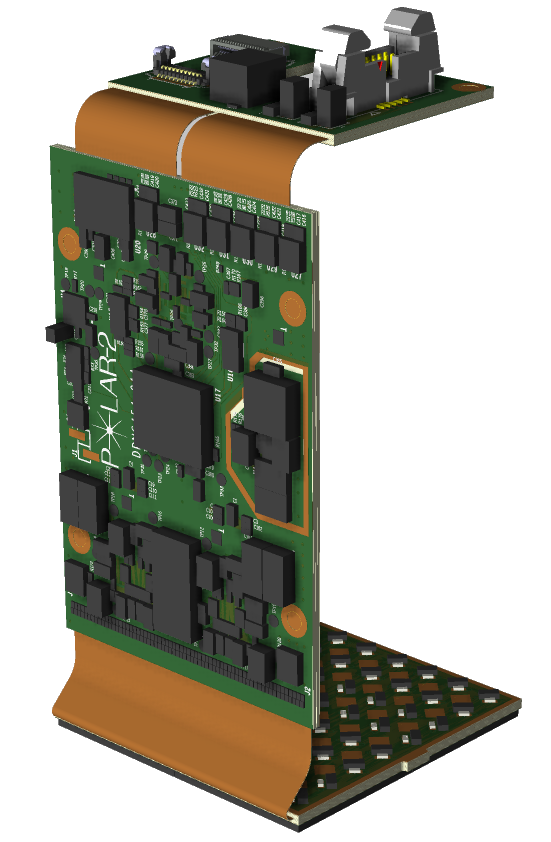}
 \caption{CAD model of the POLAR-2 instrument (left), of a single polarimeter module (middle), and of a Front-End Board (right)}
 \label{polar-2_design}
\end{figure}

The POLAR-2 instrument is made of 100 polarimeter modules, increasing by a factor 4 the number of modules compared to POLAR. The polarimeter module design, which can be seen in the middle picture of Figure \ref{polar-2_design}, has also been improved in order to increase the overall performances of the instrument. A module is composed of 64 elongated scintillator bars readout by Silicon PhotoMultiplier (SiPMs) arrays (the S13361-6075PE from Hamamatsu). The use of SiPMs increases the Photo-Detection Efficiency (PDE) by about a factor 2, thus increasing the light yield of the detector and allowing to reduce the lower energy limit of the instrument down to a few keV. Another advantage compared to MAPMTs is that the detectors can be operated at much lower voltages (several tens of Volts instead of kV). The main drawback of using SiPMs is the high dark noise, especially at high temperatures or after exposition to radiation (see \cite{SiPM_irradiation_Slawek} and \cite{SiPM_irradiation_Hiromitsu}). In order to reduce this dark current, a cooling system based on thermo-electric cooling devices (Peltier elements) has been designed. A Peltier element is placed in each polarimeter module in order to cool down the sensors and the Front-End Electronics (FEE, whose design is shown on the right picture of Figure \ref{polar-2_design}), and is connected to an external radiator through a copper threading. Thermal simulations using the Abaqus software from Dassault Systèmes\textregistered{} as well as polarimeter module-level tests \cite{POLAR-2_SPIE2020} are currently performed in order to optimize the design of this cooling system. Although a similar kind of SiPMs has already been irradiated (see \cite{SiPM_irradiation_Slawek} and \cite{SiPM_irradiation_Hiromitsu} for the irradiation of the 50$\mu$m microcells SiPMs), an irradiation campaign is planned on the S13361-6075PE arrays, as well as on some FEE components, in order to precisely characterize the behavior of such devices after being exposed to radiations.

The scintillator material is the same as that used in POLAR (EJ-248M from Eljen Technology), but the bars have been reshaped to 5.9$\times$5.9$\times$125mm$^3$. The bars are therefore wider in the polarimeter plane thanks to an optimization of the module mechanical design, reducing the dead space and therefore increasing the sensitive volume for a fixed bar length. The bar length has also been reduced from 176mm to 125mm in order to improve the Signal-to-Noise Ratio. Geant4 simulations of the instrument revealed that the optimal SNR is for 80mm long scintillators (since the background scales with the sensitive volume), and a compromise between SNR and total sensitive volume was kept in order to get high enough statistics with good quality signals.

The POLAR-2 instrument CAD design is shown on the first picture of Figure \ref{polar-2_design}. The 100 polarimeter modules (in black) are disposed on an aluminium grid (in gray). Each FEE is connected to the Back-End Electronics (BEE), which is placed in the blue part together with the power supplying system (Low Voltage Power Supply, LVPS) and the communication system to the CSS. In red is shown the adapter for the robotic arm of the station used to manipulate the payload, and in yellow is the interface plate between the payload and the platform of the CSS. The total power consumption of POLAR-2 is about 300W (2W per polarimeter module and 100W allocated to the BEE, LVPS, and communication system), while its mass and dimensions are 150kg and 590$\times$648$\times$700mm$^3$, respectively. Several spectrometer modules based on either CeBr3 or LaBr3 crystals will also be placed in POLAR-2 as a mean to increase the capability of doing spectral analysis and source position localization.

\subsection{Polarimeter Module Development and Testing}

\begin{figure}[H]
\centering
\includegraphics[height=0.38\textwidth]{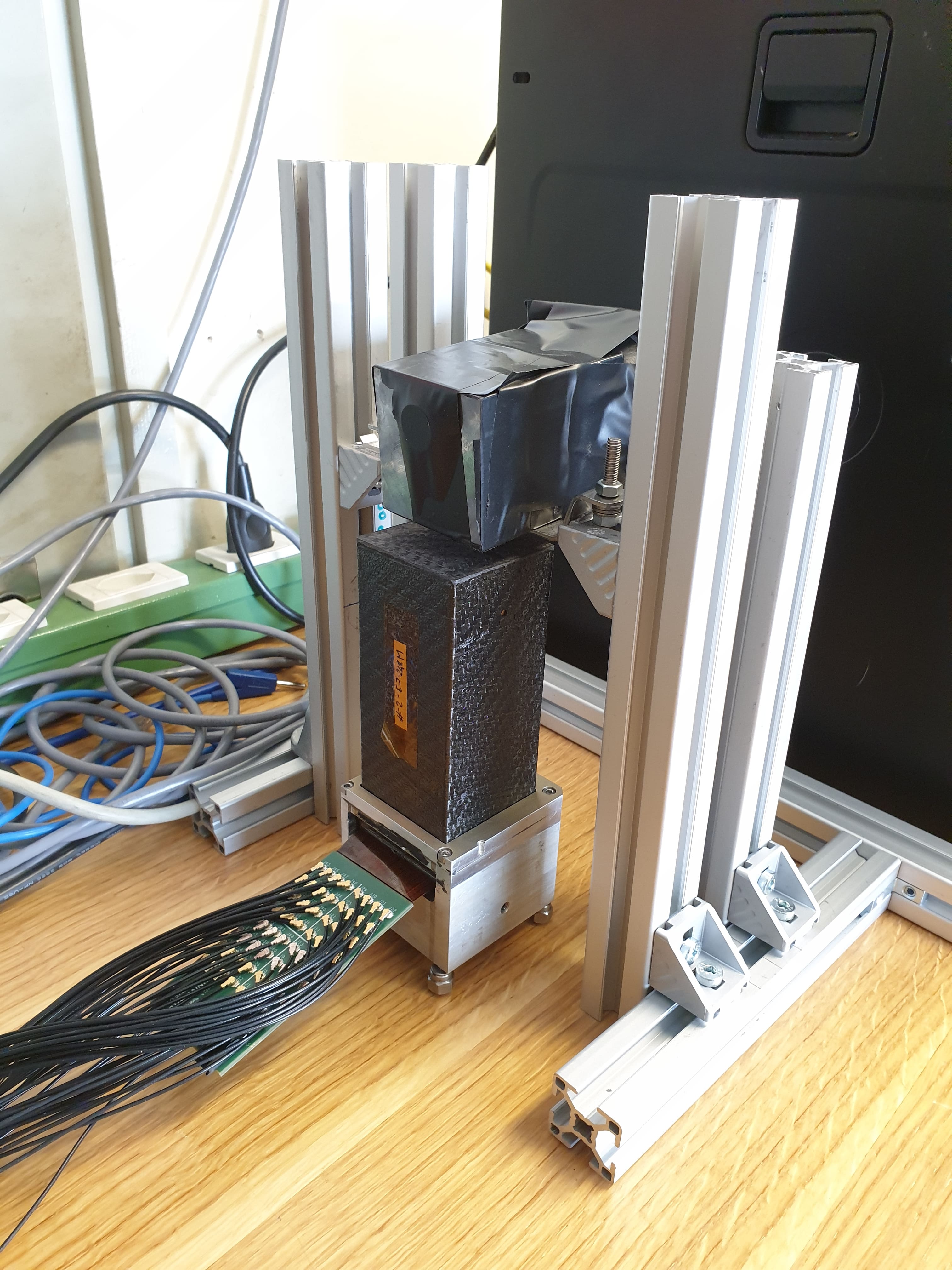}\hspace*{0.4cm}\includegraphics[height=0.38\textwidth]{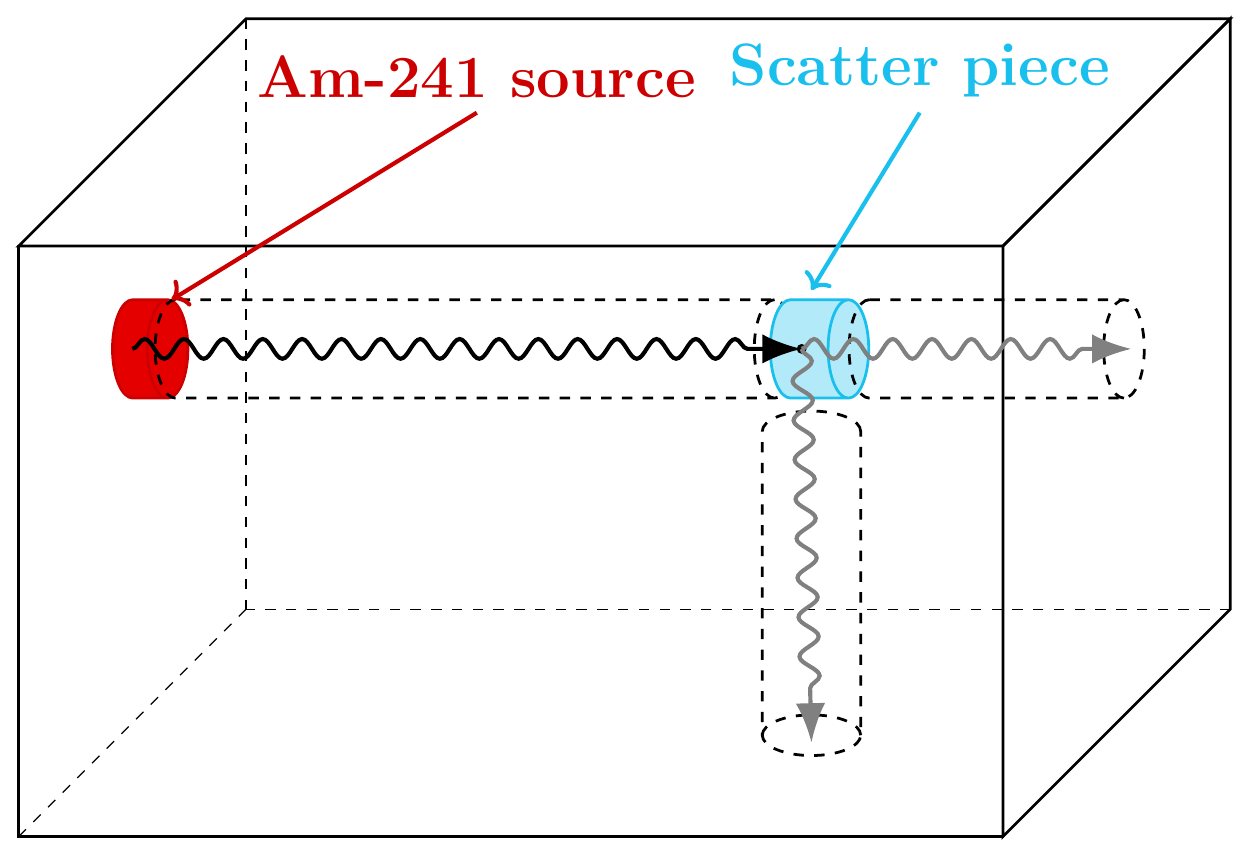}
 \caption{Module calibration setup with an $^{241}$Am source on top (left), and illustration of the method used to get polarized photons (right)}
 \label{calib_setup}
\end{figure}

Several prototype modules have been built in order to test its mechanical and optical design. The FEE mentioned previously and shown in Figure \ref{polar-2_design} has just been designed, so another readout system has been used until now in order to acquire data from the prototype modules. The SiPM arrays were readout using the front-end board of the BabyMIND experiment \cite{Babymind_board}, based on the same CITIROC-1A ASIC as will be used in the POLAR-2 front-end. A calibration setup, shown in Figure \ref{calib_setup}, has been built at CERN using a 365MBq Am241 source and a 3.37MBq Cs137. The 59.5keV photopeak of the former and the 470keV Compton edge of the latter allows to probe a wide energy range. A lead block with a plastic scatter piece has been designed in order to be able to get a polarized source of $\gamma$-rays from the Am241 source. The source can be placed on one side of this block, the photons are first focused in the direction of a plastic piece, in which they scatter. We then only select the photons scattering at 0$^\circ$ or 90$^\circ$ to get an almost 100\% polarized source with a polarization angle of 0$^\circ$ or 90$^\circ$.

\begin{figure}[H]
\centering
\hspace*{-1.1cm}\includegraphics[height=0.38\textwidth]{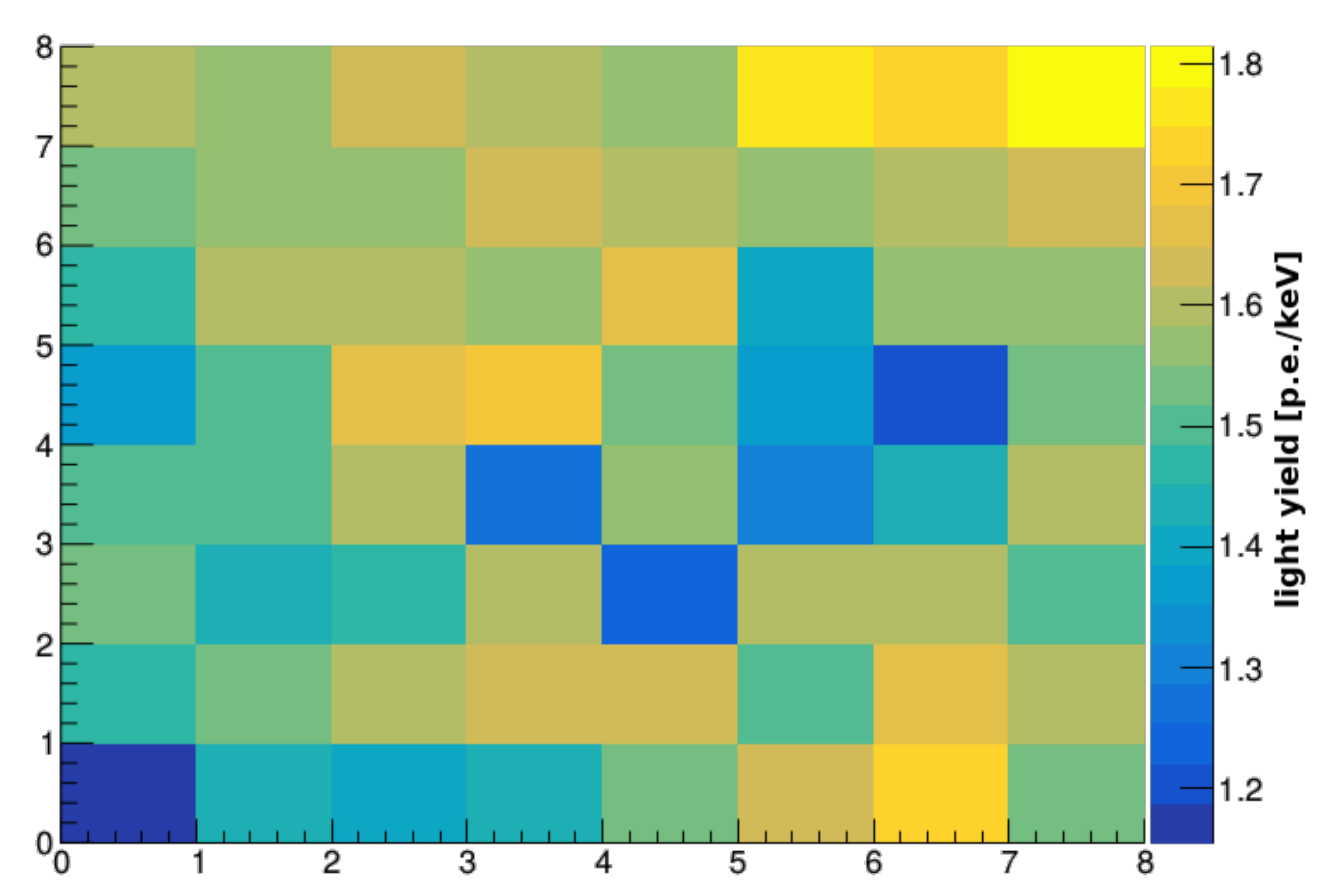}\includegraphics[height=0.38\textwidth]{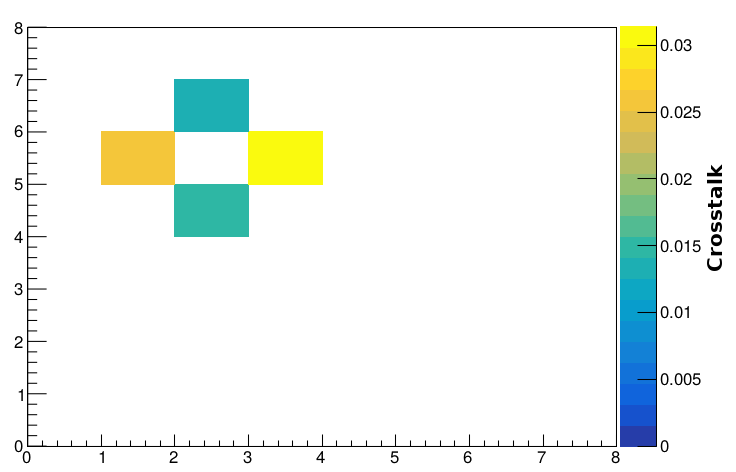}
 \caption{2D maps of the light yield in p.e./keV (left) and cross talk (right) for the $\times$8 scintillator grid inside of single POLAR-2 detector module. Some outliers can be seen in the light yield maps due to a bad alignment of the SiPM arrays w.r.t. the scintillators. In the cross talk map the cross talk from one channel to its 4 direct neighbours is shown. The cross talk to all other channels is found to be negligible.}
 \label{ly_xtalk}
\end{figure}

The light yield for each one of the 64 channels of a module is determined as follows. A single photo-electron spectrum (so called "finger plot") and the Am241 photopeak are both measured using the High Gain (HG) output of the ASIC. The finger plot allows to get the ADC to number of photo-electrons (p.e.) conversion, while the Am241 photopeak position give the ADC to keV conversion. In order to be more precise for this latter, a measurement in Low Gain (LG) is also performed with the Cs137. The Compton edge position as well as the peak separation in the finger plot lead to the light yield of the channel in p.e./keV. A map of the light yield for an entire module is shown on the left of Figure \ref{ly_xtalk}. Some outliers with lower or higher values can be seen in the map and are due to a small misalignment between the scintillators and the SiPM arrays. The misalignment issues will be mitigated using the new FEE design and a special alignment piece. The average light yield is around 1.6p.e./keV, which is over 5 times better than the 0.3p.e./keV of POLAR. This implies that we can decrease the lower bound of our energy range from 50keV to a few keV by putting a threshold at a 3 or 4 p.e.. This improvement of light yield is due to the use of SiPMs which increased the PDE and to the optimization of the mechanical design which allowed to increase the contact surface between the sensors and the scintillators.

\vspace*{-0.3cm}
Another improvement compared to POLAR is the optical cross talk from channel to channel. It went down from about 15\% in POLAR to around 1.5-2.5\%, as can be seen in the right map of Figure \ref{ly_xtalk}. Several improvements have been made in order to reduce the cross talk by an order of magnitude. The upgrade of the sensors from MAPMTs to SiPMs contributed to lower the cross talk since the MAPMTs have a single thick Borosilicate glass entrance window shared by all the anodes, whereas the SiPMs only have a thin layer of epoxy resin. The scintillator wrapping method based on Vikuiti from 3M has also been improved. Optical simulations on Geant4 \cite{Geant4} are carried out on a single module level to confirm the understanding of the observed optical behavior. The optical coupling between the SiPMs and the scintillators in the current prototype modules is made using optical grease, which cannot be used in-orbit. An optical pad based on Room Temperature Vulcanizing (RTV) silicone is currently developed. A first trial with a 0.35mm thick pad increases the optical cross talk to 4\% (since the optical pad is thicker than the layer of optical grease that was applied). Manufacturing tests are currently ongoing to decrease the optical pad thickness to below 0.2 mm, and therefore the cross talk. The cross talk values with and without optical pad are compatible with the optical simulations.\\

\vspace*{-0.6cm}
The newly designed FEE have been ordered and should be received by the end of the summer. The firmware of these boards will allow to implement the POLAR-2 trigger scheme optimized for polarization measurements. We will therefore be able to measure modulation curves and determine the $M_{100}$ (the sensitivity to polarization) of a single module. A module made with the POLAR-2 FEE will be used for a vibration and shock test next autumn at the MPE (Garching, Germany) to qualify the current mechanical design for a launch to space.

\section{Expected POLAR-2 Science Performances}

\begin{figure}[H]
\centering
\hspace*{-0.5cm}\includegraphics[height=0.4\textwidth]{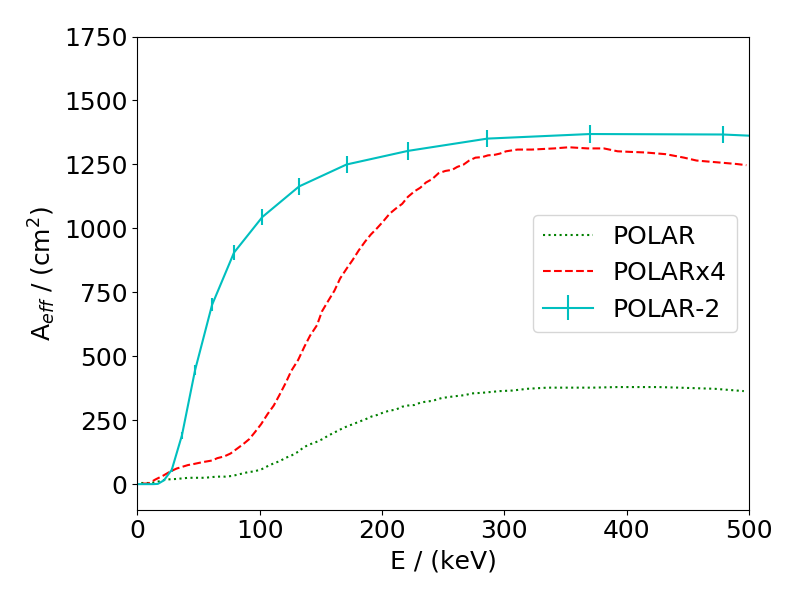}\includegraphics[height=0.4\textwidth]{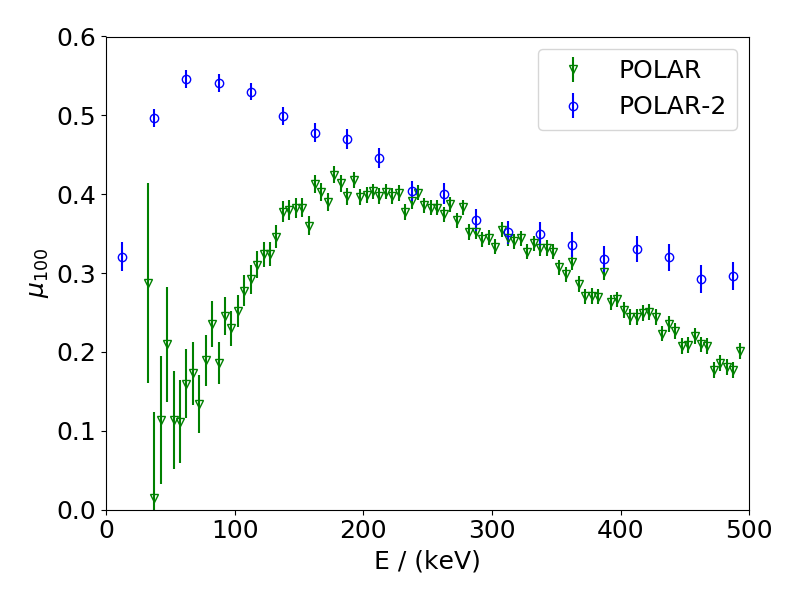}
 \caption{Effective area (left) and modulation factor $M_{100}$ (right) as a function of energy for a 26.4$^\circ$ off-axis GRB (incoming angle equal to that of GRB170114A observed by POLAR)}
 \label{aeff_mod100_polar-2}
\end{figure}

The full POLAR-2 instrument design has been implemented with the Geant4 simulation toolkit \cite{Geant4}, which allows to predict its science performances. The effective area for polarization events\footnote{A polarization event is an event that can be used for polarization studies, with therefore at least 2 triggering scintillator bars in a fixed time window (usually defined as 100~ns). Single bar events, which cannot be used for polarization analysis, will also be recorded for spectral measurements, with a higher threshold in order to limit the dead time. The effective area of POLAR-2 accounting for non-polarization events is discussed in \cite{MK_ICRC2021}.} has been simulated as a function of energy for three different configurations: the POLAR detector, a detector made of 100 POLAR-like modules (POLAR$\times$4), and the POLAR-2 detector with 100 upgraded modules. As can be seen on the left of Figure \ref{aeff_mod100_polar-2}, the gain in effective area for POLAR-2 with respect to POLAR is over an order of magnitude at low energies due to the technological improvements mentioned in the previous section. At higher energies, the POLAR-2 effective area is still above a POLAR$\times$4 detector, despite the shorter bars of POLAR-2 reducing its sensitive volume.

The modulation factor M$_{100}$, which corresponds to the relative amplitude of a 100\% polarized modulation curve, is shown for the POLAR and POLAR-2 configurations on the right side of Figure \ref{aeff_mod100_polar-2}. It can be seen that the sensitivity to polarization has clearly been improved for POLAR-2 with a peak almost reaching 0.6 at around 100~keV, whereas the POLAR M$_{100}$ did not exceed the 0.4. As discussed in \cite{MK_ICRC2021}, thanks to an improved mean minimum detectable polarization (MDP), POLAR-2 will be able to make 50 high quality GRB polarization measurements every year.



\newpage
\section*{Acknowledgements}

We gratefully acknowledge the Swiss Space Office of the State Secretariat for Education, Research and Innovation (ESA PRODEX Programme) which supported the development and production of the POLAR-2 detector. M.K. and N.D.A. acknowledge the support of the Swiss National Science Foundation. J.M.B. acknowledges support from the Alexander von Humboldt Foundation. National Centre for Nuclear Research acknowledges support from Polish National Science Center under the grant UMO-2018/30/M/ST9/00757. We gratefully acknowledge the support from the National Natural Science Foundation of China (Grant No. 11961141013, 11503028), the Xie Jialin Foundation of the Institute of High Energy Physics, Chinese Academy of Sciences (Grant No. 2019IHEPZZBS111), the Joint Research Fund in Astronomy under the cooperative agreement between the National Natural Science Foundation of China and the Chinese Academy of Sciences (Grant No. U1631242), the National Basic Research Program (973 Program) of China (Grant No. 2014CB845800), the Strategic Priority Research Program of the Chinese Academy of Sciences (Grant No. XDB23040400), and the Youth Innovation Promotion Association of Chinese Academy of Sciences.

\clearpage
\section*{Full Authors List: POLAR-2 Collaboration}


\scriptsize
\noindent
N.~De~Angelis$^1$,
J.M.~Burgess$^4$,
F.~Cadoux$^1$,
J.~Greiner$^4$,
J.~Hulsman$^1$,
M.~Kole$^1$,
H.C.~Li$^2$,
S.~Mianowski$^3$,
A.~Pollo$^3$,
N.~Produit$^2$,
D.~Rybka$^3$,
J.~Stauffer$^1$,
J.C.~Sun$^5$,
B.B.~Wu$^5$,
X.~Wu$^1$,
A.~Zadrozny$^3$
and
S.N.~Zhang$^{5,6}$ \\

\noindent
$^1$University of Geneva, DPNC, 24 Quai Ernest-Ansermet, CH-1211 Geneva, Switzerland.\\
$^2$University of Geneva, Geneva Observatory, ISDC, 16, Chemin d’Ecogia, CH-1290 Versoix, Switzerland.\\
$^3$National Centre for Nuclear Research, ul. A. Soltana 7, 05-400 Otwock, Swierk, Poland.\\
$^4$Max-Planck-Institut fur extraterrestrische Physik, Giessenbachstrasse 1, D-85748 Garching, Germany.\\
$^5$Key Laboratory of Particle Astrophysics, Institute of High Energy Physics, Chinese Academy of Sciences, Beijing 100049, China.\\
$^6$University of Chinese Academy of Sciences, Beijing 100049, China.


\begin{thebibliography}{99}

\bibitem{BATSE_catalog}
Paciesas, W. S., Meegan, C. A., Pendleton, G. N., et al. 1999, \href{https://doi.org/10.1086/313224}{\textit{ApJS}, 122, 465}

\bibitem{GW170817}
Abbott, B. P., Abbott, R., Abbott, T. D., et al. 2017, \href{https://doi.org/10.3847/2041-8213/aa920c}{\textit{ApJL}, 848, L13}

\bibitem{POLAR_design}
Produit, N., Baod, T. W., Batsch, T., et al. 2018, \href{http://dx.doi.org/10.1016/j.nima.2017.09.053}{\textit{Nucl. Instrum. Meth. Phys. Res. A}, 877, 259}

\bibitem{POLAR_NatureAstronomy}
Zhang, S. N., Kole, M., Bao, T.-W., et al. 2019, \href{http://dx.doi.org/10.1038/s41550-018-0664-0}{\textit{Nat. Astron.}, 3.3, 258}

\bibitem{POLAR_catalog}
Kole, M., De Angelis, N., Berlato, F., et al. 2020, \href{http://dx.doi.org/10.1051/0004-6361/202037915}{\textit{A\&A}, 644, A124}

\bibitem{MK_ICRC2021}
Kole, M. 2021, \href{https://pos.sissa.it/395/600/}{\textit{PoS(ICRC2021)}, 395, 600}

\bibitem{POLAR_time_resolved}
Burgess, J. M., Kole, M., Berlato, F., et al. 2019, \href{http://dx.doi.org/10.1051/0004-6361/201935056}{\textit{A\&A}, 627, A105}

\bibitem{POLAR_Crab_spectral}
Li, H. C., Gauvin, N., Ge, M.-Y., et al. 2019, \href{http://doi.org/10.1016/j.jheap.2019.10.001}{\textit{J. High Energy Astrophys.}, 24, 15}

\bibitem{HC_Crab_ICRC2021}
Li, H. C. 2021, \href{https://pos.sissa.it/395/585/}{\textit{PoS(ICRC2021)}, 395, 585}

\bibitem{SiPM_irradiation_Slawek}
Mianowski, S., Borowicz, D. M., Brylew, K., et al. 2020, \href{https://doi.org/10.1088/1748-0221/15/03/P03002}{\textit{JINST}, 15, P03002}

\bibitem{SiPM_irradiation_Hiromitsu}
Hirade, N., Takahashi, H., Uchida, N., et al., 2021, \href{https://doi.org/10.1016/j.nima.2020.164673}{\textit{Nucl. Instrum. Meth. Phys. Res. A}, 986, 164673}

\bibitem{POLAR-2_SPIE2020}
Hulsman, J. 2020, \href{https://doi.org/10.1117/12.2559374}{\textit{Proc. SPIE}, 11444}

\bibitem{Babymind_board}
Noah, E., Blondel, A., Cadoux, F. R., et  al. 2016, \href{https://doi.org/10.22323/1.252.0031}{\textit{PoS(PhotoDet2015)}, 252, 031}

\bibitem{Geant4}
Agostinelli, S., Allison, J., Amako, K., et al. 2003, \href{https://doi.org/10.1016/S0168-9002(03)01368-8}{\textit{NIM A}, 506, 250}


\end{thebibliography}
\end{document}